\documentclass[aps,preprint,a4paper,showpacs,showkeys,superscriptaddress]{revtex4-1}
\ifx\pdfoutput\undefined
\usepackage[usenames]{color} 
\usepackage[dvips]{graphicx}
\else
\usepackage{color} 
\definecolor{BrickRed}{rgb}{0.85,0.15,0.25}
\definecolor{MidnightBlue}{rgb}{0,0.45,0.85}
\definecolor{ForestGreen}{rgb}{0,0.85,0.45}
\usepackage{graphicx}
\usepackage{epstopdf}
\fi
\usepackage{subfigure}
\usepackage{latexsym,amsmath,amssymb}
\usepackage{hyperref}     
\hypersetup{colorlinks,%
  citecolor=blue,%
  linkcolor=cyan,%
  pdftex}
\allowdisplaybreaks

\usepackage{umoline}
\newsavebox\CBox

\begin{document}


\title{Hawking Effect of AdS$_2$ Black Holes in the Jackiw-Teitelboim Model}

\author{Wontae Kim}%
\email[]{wtkim@sogang.ac.kr}%
\affiliation{Department of Physics, Sogang University, Seoul 121-742,
  South Korea}%

\date{\today}

\begin{abstract}
It might be tempting to consider that
the two-dimensional anti-de Sitter black hole in the Jackiw-Teitelboim model
is thermally hot by invoking the non-vanishing surface gravity, which raises a natural question of ``where is the observer to measure the temperature?",
asymptotically at infinity or in a finite region outside the horizon?
In connection with this issue, one might expect that the local temperature
would also be blue-shifted near the horizon while it would vanish at infinity
because of the Tolman factor in the local temperature.
In this paper, the local temperature will be shown to vanish and to respect the equivalence principle everywhere as long as a consistent Stefan-Boltzmann law is required. The essential reason for the vanishing local temperature
will be discussed on various grounds.

\end{abstract}



\maketitle


\begin{centering}
\section{Introduction}
\label{sec:intro}
\end{centering}
The thermal property of a Schwarzschild black hole
has been characterized by the Hawking temperature
at infinity
 \cite{Hawking:1974sw}.
In the Israel-Hartle-Hawking state
\cite{Israel:1976ur,Hartle:1976tp},
one can also define the local Tolman temperature
derived from  the Stefan-Boltzmann law
to relate the local energy density and the local temperature
in local proper frames \cite{Tolman:1930zza,Tolman:1930ona}.

Let us consider a two-dimensional black hole for a simple argument. Then,
the local temperature of the black hole can be obtained from the Stefan-Boltzmann law
\begin{equation}
\rho = \gamma T^{2}=T_{\rm H}/\sqrt {g}, \label{SB1}
\end{equation}
where $\rho$ is the local energy density and
$T$ is the Tolman temperature, which consists of the Hawking temperature $T_{\rm H}$ and the redshift factor.
The issues are as follows:
Firstly, in thermal equilibrium, the local energy density is generally
negative finite near the horizon up to a finite distance \cite{Visser:1996iw};
thus, the Tolman temperature is plagued by imaginary values in that region as long as
the usual Stefan-Boltzmann law is employed.
This negative energy density also appears in the same manner
in the four-dimensional Schwarzschild black hole \cite{Page:1982fm}.
Secondly, the Tolman temperature is divergent at the horizon
despite the finite local energy density, which means that the usual form of the
Stefan-Boltzmann law is inconsistent \cite{Gim:2015era}.
Thirdly, in the Israel-Hartle-Hawking state of the equilibrium system,
there are no effective excitations at the horizon
so that the equivalence principle
can be naturally restored there \cite{Singleton:2011vh,Gim:2015era,Barbado:2016nfy},
which implies that the thermal temperature must vanish on the horizon.

To resolve these issues, one should note that the Tolman temperature from
the Stefan-Boltzmann law has been derived
from the assumption that the energy-momentum tensor
is traceless \cite{Tolman:1930zza,Tolman:1930ona}.
Hence, the usual Stefan-Boltzmann law should be extended to the anomalous case
of an energy-momentum tensor with a non-vanishing trace,
where the Hawking process of radiation
in the black hole systems is actually responsible for the conformal anomaly for matter fields \cite{Christensen:1977jc}.
For an energy-momentum tensor with a non-vanishing trace,
one can show that the local temperature is effectively zero
at the horizon so that the above issues can be naturally resolved
for two and four-dimensional Schwarzschild black holes \cite{Gim:2015era,Eune:2015xvx}
and for the Schwarzschild anti-de Sitter (AdS) black hole \cite{Eune:2017iab}.
(For a recent review, see Ref.~\cite{Kim:2017aag}.)

However, this still raises a question on the behavior of the local temperature of
an AdS black hole
with a constant curvature. The essential difference
from the above asymptotically flat black holes is that
the local temperature for AdS black holes does not reduce to the Hawking temperature
at infinity because of the AdS boundary. Moreover, the AdS geometry is locally
equivalent and it is indistinguishable.
Regarding the calculations of the local temperature for the AdS black hole,
one may use the local temperature identified with
the Unruh temperature for an accelerating observer in a higher-dimensional
Minkowski spacetime~\cite{Deser:1997ri, Deser:1998xb,Brynjolfsson:2008uc};
however, the local temperature suffers from an imaginary value.

Thus, we would like to consider an amenable two-dimensional AdS black hole
with a constant curvature in order to study whether
the local temperature of the AdS black hole can be well-defined without imaginary values
and eventually justify whether the black hole is thermally hot or not.
In Section \ref{JT}, we consider a soluble AdS$_2$ black hole in
the Jackiw-Teitelboim (JT) model \cite{Teitelboim:1983ux,Jackiw:1984je} and
find a clue to the behavior of
the local temperature from the fact that
the general covariance requires
a vanishing energy-momentum tensor related to
the absence of the Hawking flux.
This implies that the local temperature vanishes everywhere.
In order to confirm this result, in Section~\ref{sec:Tolman},
we derive the local temperature in the presence of an
energy-momentum tensor with a non-trivial trace. We will show that the
effective local temperature for the AdS$_2$ black hole becomes zero without any imaginary
values.
In Section \ref{sec:excitations}, the claims in the Section~\ref{sec:Tolman}
will be clarified by pointing out some differences from the result of an earlier work \cite{Spradlin:1999bn}.
Finally, a conclusion will be given in Section \ref{conclusion}.

\begin{centering}
\section{AdS$_2$ black hole in the JT model}
\label{JT}
\end{centering}
Let us start with the JT model described by the action
\cite{Teitelboim:1983ux,Jackiw:1984je}
\begin{equation}
S_{\rm JT}=\frac{1}{2 \pi}\int d^2 x \sqrt{-g}\Phi \left[R +\frac{2}{\ell^2} \right],
\end{equation}
where $\Phi$ is an auxiliary field. This model has been extensively studied  from various points of view \cite{Mann:1989gh,Muta:1992xw,Lemos:1993qn,Kumar:1994ve,Lemos:1996bq}.
The actions for classical and quantum matter
can be written in the forms of
\begin{eqnarray}
S_{{\rm{CL}}} &=&  \frac{1}{2 \pi} \int\/d^{2}x \sqrt{-g}\,
       \left[ - \frac12  (\nabla f)^2 \right],
\label{cm}\\
S_{{\rm{QT}}} &=& -\frac{1}{96\pi}\int d^2x\/ \sqrt{-g}\left[
         R\frac{1}{\Box}R \right],
\label{qm}
\end{eqnarray}
where $f$ is a classical scalar field and Eq. \eqref{qm} is the Polyakov action \cite{Polyakov:1987zb}.
In the conformal gauge
\begin{equation}
ds^2=- e^{2\rho(\sigma^+,\sigma^-)} d\sigma^+ d\sigma^-, \label{metric1}
\end{equation}
the equations of motion for $\rho$, $\Phi$, and $f$, and the constraint equations
are given as
\begin{eqnarray}
&& \partial_+\partial_- \rho + \frac{1}{4\ell^2} e^{2\rho}=0,
            \label{dynamic} \\
&&\partial_+\partial_- \Phi-\frac{1}{4l^2}e^{2\rho}(\frac{1}{24}-\Phi)=0,\\
&& \partial_+\partial_- f=0,\label{f} \\
&&\partial_\pm^2\Phi - 2\partial_\pm \rho \partial_\pm \Phi=T_{\pm\pm}^{\rm M}.
 \label{jtcon}
\end{eqnarray}
The energy-momentum tensor for matter fields consists of two parts:
\begin{equation}
T_{\pm \pm}^{\rm M}=T_{\pm \pm}^{\rm CL}+T_{\pm \pm}^{\rm QT},
\end{equation}
where
\begin{eqnarray}
T_{\pm\pm}^{\rm CL}&=&\frac12\left(\partial_\pm f\right)^2, \\
T_{\pm \pm}^{\rm QT} &=&T_{\pm\pm}^{\rm bulk}+T_{\pm\pm}^{\rm boundary}, \label{boundary1}
\end{eqnarray}
with $\kappa=1/(12\pi)$. For later convenience, the  energy-momentum tensor $T_{\pm \pm}^{\rm QT}$ can be
divided into two parts as
\begin{eqnarray}
\label{bulkya}
T_{\pm\pm}^{\rm bulk} &=& - \kappa \left[ \left(\partial_\pm\rho\right)^2- \partial_\pm^2\rho\right],
\end{eqnarray}
and the boundary term becomes
\begin{equation}
T_{\pm\pm}^{\rm boundary}= -\kappa t_\pm \label{boun}
\end{equation}
where $t_{\pm}$ reflects the nonlocality of the Polyakov action in Eq.~\eqref{qm}.
Solving the equations of motion
in the conformal gauge, we obtain
\begin{eqnarray}
& & e^{2\rho}=\frac{M}{{\rm sinh}^2\left[\frac{\sqrt{M}(\sigma^+ -\sigma^-)}
                {2\ell} \right]},  \label{metric} \\
& & \Phi^{-1}=-\frac{1}{M} {\rm tanh}\left[\frac{\sqrt{M}(\sigma^+ -\sigma^-)}
                {2\ell} \right]            ,\\
& & f_i =f_i^{(+)}(\sigma^+) +f_i^{(-)}(\sigma^-) \label{matsol},
\end{eqnarray}
where
$M $ is an integration constant that plays the role of the mass of the black hole.
The curvature scalar related to the trace of the energy-momentum tensor
${T^{\rm QT}}^{\mu}_{\mu} = R/(24\pi)$ is given as
\begin{equation}
R= -\frac{2}{\ell^2},
\end{equation}
which is independent of the mass parameter.
As $M \rightarrow 0$, the solution \eqref{metric} exactly comes down to
the AdS$_2$ vacuum
\begin{equation}
\label{vacuum}
e^{2\rho}=\frac{4\ell^2}{(x^+ - x^-)^2}.
\end{equation}
In fact, the black hole and the vacuum geometry described by Eqs.~\eqref{metric} and \eqref{vacuum}, respectively,
are related through the
coordinate transformation
\begin{eqnarray}
x^\pm =\frac{2\ell}{\sqrt{M}} {\rm tanh}
\frac{\sqrt{M}\sigma^\pm}{2\ell}. \label{sigmaandy}
\end{eqnarray}
Furthermore, the dynamical equation of motion
\eqref{dynamic} is decoupled from the matter field and
the constant curvature is independent of the energy density of infalling matter, so we set  $f =0$ from now on.

Let us now consider the general coordinate transformation for the energy-momentum tensor
in order to calculate the Hawking flux for the AdS$_2$ black
hole in the JT model.
The Hawking radiation in two dimensions
can be given by the anomalous transformation
of the boundary term in the energy-momentum tensor.
We assume that the constraint equations (\ref{jtcon}) should be Virasoro
anomaly free in such a way that $T_{\pm\pm}^{\rm QT}$ can be transformed as
the primary
operator in conformal field theory as \cite{Bilal:1992kv,NavarroSalas:1995vd,Cadoni:1994uf,Kim:1995wr}
\begin{equation}
T_{\pm\pm}^{\rm QT}(\sigma^\pm)= \left(\frac{\partial_\pm x^\pm}{\partial_\pm
  \sigma^\pm}\right)^2T_{\pm\pm}^{\rm QT}(x^\pm);
\end{equation}
then, the boundary term in Eq.~ \eqref{boundary1} is determined by
\begin{eqnarray}
\label{sch}
-t_{\pm}(\sigma^{\pm})&=&-\frac{1}{2} \{x^{\pm}, \sigma^{\pm} \} \nonumber \\
                      &=&\frac{M}{4  \ell^2},
\end{eqnarray}
where $\{x^-, \sigma^- \}$ is a Schwartzian derivative.
At first sight,
one might conclude that the Hawking temperature is determined by
the ordinary relation $- t_-= \pi^2 T_H^2$  in Refs. \cite{Bilal:1992kv,NavarroSalas:1995vd,Kim:1995wr} and becomes
\begin{equation}
T_{H}=\frac{\sqrt{M}}{2\pi \ell}. \label{HT}
\end{equation}
However, this is not so for the asymptotically non-flat case
because the bulk and the boundary terms contribute to the Hawking radiation
simultaneously unlike in the asymptotically flat case,
\begin{eqnarray}
T_{--}^{\rm bulk}
= -\frac{\kappa M}{4\ell^2},\label{bulk2}~~ T_{--}^{\rm boundary}= \frac{\kappa M }{4\ell^2};
\label{boundary2}
\end{eqnarray}
thus, the radiation $h$
vanishes as
\begin{eqnarray}
\label{null}
h(\sigma^+,\sigma^-) &=&T_{--}^{\rm QT} (\sigma^+,\sigma^-) \nonumber \\
                     &=& T_{--}^{\rm bulk}+ T_{--}^{\rm boundary}
                                  \nonumber \\
                      &=& 0.
\end{eqnarray}
The negative contribution from the bulk part of Eq.~\eqref{bulk2}
is in contrast with
the asymptotically flat case, for instance, a CGHS black hole that is asymptotically flat \cite{Callan:1992rs}.
Explicitly, the Hawking radiation for the CGHS black hole
is given by
\begin{eqnarray} \label{cghs}
h(\sigma^-)&=&T_{--}^{\rm QT}(\sigma^+,\sigma^-)
                  |_{\sigma^+ \rightarrow \infty}\nonumber \\
          &=& T_{--}^{\rm bulk}|_{\sigma^+ \rightarrow \infty}+
            T_{--}^{\rm boundary}|_{\sigma^+ \rightarrow \infty} \nonumber \\
     &=& -\kappa t_- (\sigma^-).
\end{eqnarray}
In the limit of asymptotic null infinity $(\sigma^+ \rightarrow + \infty)$,
the Hawking radiation originates only from the
boundary term $t_-$
because the bulk contribution
vanishes at null infinity and the Stefan-Boltzmann law can be
well-defined as $- t_-= \pi^2 T_H^2$ at infinity.
In this respect, we can also expect
the local temperature in the proper frame to be zero for the AdS$_2$ black hole,
which will be clarified by calculating it explicitly
in the next section.\\

\begin{centering}
\section{Effective Local temperature}
\label{sec:Tolman}
\end{centering}
The argument in the previous section will be elaborated by calculating
the local temperature directly. For this purpose, we
recapitulate the derivation of the
Stefan-Boltzmann law in order to figure out the local temperature
for the AdS black hole where there is no asymptotic flat region. As a matter of fact,
the conventional local temperature relying on the Tolman's form is based on
the traceless energy-momentum tensor.
However, the energy-momentum tensor of matter fields is anomalous
because of the trace anomaly \cite{Deser:1976yx}, so
the usual Tolman temperature should be generalized to the
case of a non-vanishing trace because the trace of
the energy-momentum tensor is, indeed,
non-vanishing for an AdS$_2$ black hole \cite{Christensen:1977jc}.

Let us start with the first law of thermodynamics, which is assumed to be
$dU =TdS-pdV$,
where $U$, $T$, $S$, and $V$ are the thermodynamic internal energy, temperature, entropy, and
volume in the proper frame, respectively, and $U=\int \rho dV$.
Using the Maxwell relation $\partial S/\partial V \vert_T = \partial p/\partial T\vert_V$,
we can get the first-order differential equation for the energy density as
\begin{equation}
\label{key}
2\rho = T \left.\frac{\partial \rho}{\partial T}\right\vert_V - T^\mu_\mu,
\end{equation}
where we have used the fact that
the conformal anomaly is independent of the temperature \cite{BoschiFilho:1991xz}.
Solving Eq. \eqref{key}, we obtain the proper energy density as \cite{Gim:2015era}
\begin{align}
\label{r}
\rho = \gamma T^2 - \frac{1}{2}T^\mu_\mu,
\end{align}
which reduces to the conventional Stefan-Boltzmann form for $T^\mu_\mu=0$.
In this case, the two-dimensional Stefan-Boltzmann constant is identified with $\gamma= \pi/6$
for a massless scalar field \cite{Christensen:1977jc}.

Let us perform the coordinate transformation
from tortoise coordinates to Schwarzschild coordinates
by using the relation $r^* =-(\ell/\sqrt{M}) {\rm tanh^{-1}}
r /(\ell \sqrt{M})$ in order to specify the static local observer easily.
Then, the two-dimensional line element
\eqref{metric1} can be rewritten as
\begin{eqnarray}\label{metric2}
ds^2=- g(r)dt^2+\frac{1}{g(r)} dr^2,
\end{eqnarray}
where $g(r)=-M +(r/\ell)^2$.
The horizon of the black hole is $r_{\rm H}=\sqrt{M}\ell$, and
the Hawking temperature \eqref{HT} is given by the surface gravity.
The one-loop quantum-mechanical energy-momentum tensors \eqref{qm} can be rewritten
in the static coordinates as
\begin{align}
 T_{\pm\pm}^{\rm QT}  &=\frac{1}{96\pi}\left[gg''-\frac{1}{2}(g')^2 -8t_\pm  \right]=-\frac{\kappa M}
 {4 \ell^2} - \kappa t_{\pm}, \label{flux} \\
 T_{+-}^{\rm QT} &=\frac{\kappa}{4}gg'' = \frac{\kappa}{4\ell^2} g.
\end{align}
In the Israel-Hartle-Hawking state \cite{Israel:1976ur,Hartle:1976tp},
there are no inward and outward fluxes at the past and the future
event horizon, so the boundary condition should be determined by
\begin{equation}
-t_{\pm}=\frac{M}{4 \ell^2}. \label{boundary}
\end{equation}
Note that the Boulware state
is the state appropriate to the vacuum around a static gravitational systems and contains
no radiation at spatial infinity \cite{Boulware:1974dm}; however, the state is indistinguishable from either Israel-Hartle-Hawking state or the Unruh state \cite{Unruh:1977ga} because the energy-momentum tensor \eqref{flux} has no spatial dependence
in the static Schwarzschild-like coordinates so that three vacua share the same boundary condition \eqref{boundary}.
This feature is in contrast to the asymptotically flat black holes
in that all the vacuum states are degenerate for the AdS$_2$ case.
The absence of the Unruh vacuum has something to do with the stability of the AdS$_2$ black hole
without evaporation of the black hole \cite{Kim:1998wy}.
In fact, if one wanted to consider an evaporating black hole, the boundary would be
required to become absorptive by allowing energy to be transferred between bulk fields and
external fields \cite{Rocha:2008fe}.

For the proper frame dropped from rest, the velocity from the geodesic equation of motion
can be written as
\begin{equation}\label{velocity}
u^\mu=\frac{dx^\mu}{d\tau} = \left(\frac{1}{ \sqrt{g(r)}},~~0\right),
\end{equation}
and the local energy density is given as
\begin{eqnarray}
\rho &=&T_{\mu\nu}^{\rm QT}u^\mu u^\nu \label{localenergydensity} \\
      &=& \frac{1}{g} \left[ T_{++}^{\rm QT} +T_{--}^{\rm QT} + 2 T_{+-}^{\rm QT} \right]. \label{energydensity}
\end{eqnarray}
Plugging the energy density \eqref{energydensity} into Eq. \eqref{r},
we finally obtain
\begin{equation}
\gamma T^2 = \frac1g \left[ T_{++}^{\rm QT} +T_{--}^{\rm QT}\right]=0. \label{localtem}
\end{equation}
Interestingly the effective local temperature \eqref{localtem} should vanish
generically because the inward and the outward fluxes crossing the horizon do not exists at the horizon.
In fact, we treated the appropriate limiting procedure because the denominator in Eq.
\eqref{localtem} also
vanishes on the horizon. As a result, for the
AdS$_2$ case, from Eqs. \eqref{flux} and \eqref{boundary},
the local temperature \eqref{localtem} naturally vanishes everywhere,
and the equivalence principle turns out to be perfectly valid
as compared to the asymptotically flat black hole, which satisfies the equivalence
principle just on the horizon \cite{Singleton:2011vh}.
If one were to insist on using the conventional Tolman temperature of $\rho = \gamma T^{2}=T_{\rm H}/\sqrt {g}$
rather than Eq. \eqref{r},
then the fluxes should be defined as $T_{\pm\pm}^{\rm QT}=M \kappa/4\ell^2$
for $ t_{\pm}=M/2  \ell^2 $. Unfortunately,
the Tolman temperature for such a case is divergent in spite of the
finite local energy density at the horizon, which is inconsistent.\\

\begin{centering}
\section{Particle excitations and Temperature}
\label{sec:excitations}
\end{centering}
One might wonder the reason for the claims in the previous section looking different from those in an earlier work. Now, we
are going to explain some differences between our claims and the results based on previous claims
by using the following alternative method: at first sight, the Hawking radiation in the
AdS$_2$ black hole is analogous to that in the Rindler case, because the black-hole
geometry \eqref{metric} can be obtained through a coordinate transformation from
the AdS$_2$ vacuum \eqref{vacuum} where the energy-momentum tensor is always zero. However, if
we employ suitable coordinates \eqref{sigmaandy}, we will observe particle
excitations \eqref{HT}, and we would conclude that the state is thermal \cite{Spradlin:1999bn}.
However, this is not the case for the two-dimensional AdS black hole.

Let us first consider the Unruh effect \cite{Unruh:1983ms},
where the particles are excited in the Rindler spacetime.
The normal ordered energy-momentum tensor in the Minkowski spacetime
is defined by
\begin{equation}\label{vev}
:T_{\mu\nu}:=T_{\mu\nu}-~ _{\rm M}\langle0| T_{\mu\nu} |0\rangle _{\rm M},
\end{equation}
where the
vacuum expectation value of the energy-momentum tensor gives rise to
the quadratic divergence in
two dimensions.
In general, the normal ordering depending on the coordinates
breaks the tensorial property of the energy-momentum tensor in such a way that
the normal ordered one does not transform as a tensor
under coordinate transformations:
\begin{equation}
:T_{\pm\pm}(y^{\pm}):=\left(\frac{\partial x^{\pm}}{\partial y^{\pm}}\right)^2 :T_{\pm\pm}(x^{\pm}):-\frac{\kappa}{2} \{x^{\pm},y^{\pm}\}. \label{transformation}
\end{equation}
The anomalous term comes from the coordinate transformation of the
vacuum expectation value of the energy-momentum tensor in Eq. \eqref{vev}.
Note that
the Schwartzian derivative is innocuous because the coordinate transformation is global.
For a linear Lorentz transformation,
the energy-momentum tensor is a true tensor because the Schwartzian derivative vanishes.
In particular, for the Rindler coordinate transformation of $x^\pm =\pm \left(1/a\right) e^{\pm a y^\pm}$, where $x^{\pm}$ and $y^{\pm}$ are the Minkowski and the Rindler coordinates, respectively,
the vacuum expectation value of the energy-momentum tensor in Rindler coordinates is
\begin{equation}
_{\rm M}\langle 0| :T_{\pm\pm}(y^\pm): |0\rangle _{\rm M} = \frac{\kappa}{4} a^2,
\end{equation}
where $\{x^{\pm},y^{\pm}\}=-a^2/2$, with $a$ being a linear acceleration, and $|0\rangle _{\rm M}$ is the ordinary Minkowski vacuum.
Using $\rho=\gamma T^2$ with $\gamma=\pi/6$ for a single scalar field,
where $\rho=T^{00}=\langle T_{++}\rangle+\langle T_{--}\rangle+2\langle T_{+-}\rangle$
with $\langle T_{+-}\rangle=0$ in the Rindler space,
the Unruh temperature can be obtained as
\begin{equation}
T=\frac{a}{2\pi}.
\end{equation}
From this non-tensorial quantity,
we obtained the Unruh temperature of the accelerated detector, but it is still physical
because the coordinate transformation is global, so general covariance is not
involved.

On the other hand,
the above calculations can be extended to curved spacetime by adding the bulk contribution  to the normal ordered
energy-momentum tensor \cite{doi:10.1142/p378}:
\begin{equation}
T_{\mu\nu}=T_{\mu\nu}^{\rm bulk} + :T_{\mu\nu}:, \label{realtensor}
\end{equation}
where $T_{\pm\pm}^{\rm bulk} =
- \kappa \left[ \left(\partial_\pm\rho\right)^2-
 \partial_\pm^2\rho\right]$ and $T_{+-}^{\rm bulk} =-\kappa \partial_+ \partial_- \rho$
 in the conformal gauge \eqref{metric1}. The resulting
energy-momentum tensor consists of the vacuum polarization of the
scalar field on the curved spacetime and the normal-ordering effect associated with
the coordinate system. The conformal coordinate transformation of the bulk part is calculated as
\begin{eqnarray}
T_{\pm\pm}^{\rm bulk}(y^\pm) & = & \left(\frac{\partial x^\pm}{\partial y^\pm}\right)^2 T_{\pm\pm}^{\rm bulk} (x^\pm)+\frac{\kappa}{2}\{x^{\pm},y^{\pm}\} \label{tensor2}
\end{eqnarray}
while the coordinate transformation of
the normal ordered energy-momentum tensor is implemented by using Eq. \eqref{transformation}.
Because the  Schwartzian derivatives in Eqs. \eqref{tensor2} and \eqref{transformation} exactly cancel, the energy-momentum tensor \eqref{realtensor} behaves as a true tensor
under the coordinate transformation \cite{NavarroSalas:1995vd}:
\begin{equation}
T_{\pm\pm}(y^{\pm})=\left(\frac{\partial x^{\pm}}{\partial y^{\pm}}\right)^2 T_{\pm\pm}(x^{\pm});
\end{equation}
in other words, it is generally covariant.
Then, the boundary term \eqref{boun}
can be identified with the normal ordered part in Eq. \eqref{realtensor}, so
$T_{\pm\pm}^{\rm boundary}= :T_{\pm\pm}:=-\kappa t_{\pm}$ \cite{doi:10.1142/p378}.

Unlike the Rindler case, in order for general covariance, the contribution of the bulk energy-momentum tensor
should no longer be ignored in
curved spacetime. Therefore,
the physical temperature should be derived in the regime of the generally covariant framework
when gravity couples to matter, which is a big difference from the
derivation of the Unruh temperature. This is also the reason why the bulk contribution could not be ignored
in Section \ref{sec:Tolman}. Of course, a generally covariant quantity such as the
energy-momentum tensor cannot be translated directly to the physical temperature because
it is still coordinate dependent, except at asymptotic infinity. For an
asymptotically flat black hole such as a CGHS black hole,
the bulk contribution can be safely ignored, as seen from Eq.
\eqref{cghs}, so the procedure to get the temperature looks similar to that for the Unruh case.
However, our concern is for the AdS$_2$ black hole without any other flat regions.
In that sense, considering a local quantity defined in a
local inertial frame, such as
$\rho =T_{\mu\nu}^{\rm QT}u^\mu u^\nu$ already defined in Eq. \eqref{localenergydensity}, is natural.

\begin{centering}
\section{Conclusion}
\label{conclusion}
\end{centering}
The local temperature for the AdS$_2$ black hole in the JT model was shown to vanish everywhere, which respects the equivalence principle
everywhere.
If the local temperature
of the black hole is identified with the usual Tolman temperature,
then the usual Tolman temperature is incompatible with
the Stefan-Boltzmann law defined in the presence of the conformal anomaly.
Furthermore, the three vacua of the
Boulware, Israel-Hartle-Hawking, and Unruh states were found to be degenerate in the JT model so that the black hole is stable
without any evaporation.

Let us comment on the behaviors of the local temperatures
of asymptotically flat and non-flat black holes.
For the Schwarzschild black hole,
the equivalence principle could be recovered
just at the horizon
because the temperature measured by a fixed observer in
the gravitational background is generically higher than the Unruh temperature of an accelerating observer;
however, they are the same at the event horizon of the black hole so that the equivalence principle in the
quantized theory is restored at the horizon \cite{Singleton:2011vh}.
In other words, the local temperature should vanish at the horizon and excited quanta cannot be found on the horizon,
which is compatible with the calculations of
the local temperature \cite{Gim:2015era,Eune:2015xvx}.
On the other hand, in the AdS$_2$ black hole,
the influx and the outward fluxes are zeros from the boundary conditions so that
their coordinate transformation into the local inertial frame is also trivial and
the corresponding local temperature is in essence zero.

\acknowledgments
I would like to thank Eune for discussions.
This work was supported by the National Research Foundation of Korea (NRF) grant funded by the
Korea government (MSIP) (2017R1A2B2006159).


\bibliographystyle{JHEP}       

\bibliography{references}

\providecommand{\href}[2]{#2}\begingroup\raggedright\begin{thebibliography}{10}

\bibitem{Hawking:1974sw}
S.~W. Hawking, \emph{{Particle Creation by Black Holes}},
  \href{http://dx.doi.org/10.1007/BF02345020}{\emph{Commun. Math. Phys.} {\bf
  43} (1975) 199--220}.

\bibitem{Israel:1976ur}
W.~Israel, \emph{{Thermo field dynamics of black holes}},
  \href{http://dx.doi.org/10.1016/0375-9601(76)90178-X}{\emph{Phys. Lett.} {\bf
  A57} (1976) 107--110}.

\bibitem{Hartle:1976tp}
J.~B. Hartle and S.~W. Hawking, \emph{{Path Integral Derivation of Black Hole
  Radiance}}, \href{http://dx.doi.org/10.1103/PhysRevD.13.2188}{\emph{Phys.
  Rev.} {\bf D13} (1976) 2188--2203}.

\bibitem{Tolman:1930zza}
R.~C. Tolman, \emph{{On the Weight of Heat and Thermal Equilibrium in General
  Relativity}}, \href{http://dx.doi.org/10.1103/PhysRev.35.904}{\emph{Phys.
  Rev.} {\bf 35} (1930) 904--924}.

\bibitem{Tolman:1930ona}
R.~Tolman and P.~Ehrenfest, \emph{{Temperature Equilibrium in a Static
  Gravitational Field}},
  \href{http://dx.doi.org/10.1103/PhysRev.36.1791}{\emph{Phys. Rev.} {\bf 36}
  (1930) 1791--1798}.

\bibitem{Visser:1996iw}
M.~Visser, \emph{{Gravitational vacuum polarization. 1: Energy conditions in
  the Hartle-Hawking vacuum}},
  \href{http://dx.doi.org/10.1103/PhysRevD.54.5103}{\emph{Phys. Rev.} {\bf D54}
  (1996) 5103--5115}, [\href{http://arxiv.org/abs/gr-qc/9604007}{{\tt
  gr-qc/9604007}}].

\bibitem{Page:1982fm}
D.~N. Page, \emph{{Thermal Stress Tensors in Static Einstein Spaces}},
  \href{http://dx.doi.org/10.1103/PhysRevD.25.1499}{\emph{Phys. Rev.} {\bf D25}
  (1982) 1499}.

\bibitem{Gim:2015era}
Y.~Gim and W.~Kim, \emph{{A Quantal Tolman Temperature}},
  \href{http://dx.doi.org/10.1140/epjc/s10052-015-3765-2}{\emph{Eur. Phys. J.}
  {\bf C75} (2015) 549}, [\href{http://arxiv.org/abs/1508.00312}{{\tt
  1508.00312}}].

\bibitem{Singleton:2011vh}
D.~Singleton and S.~Wilburn, \emph{{Hawking radiation, Unruh radiation and the
  equivalence principle}},
  \href{http://dx.doi.org/10.1103/PhysRevLett.107.081102}{\emph{Phys. Rev.
  Lett.} {\bf 107} (2011) 081102}, [\href{http://arxiv.org/abs/1102.5564}{{\tt
  1102.5564}}].

\bibitem{Barbado:2016nfy}
L.~C. Barbado, B.~Carlos, L.~J. Garay and G.~Jannes, \emph{{Hawking versus
  Unruh effects, or the difficulty of slowly crossing a black hole horizon}},
  \href{http://dx.doi.org/10.1007/JHEP10(2016)161}{\emph{JHEP} {\bf 10} (2016)
  161}, [\href{http://arxiv.org/abs/1608.02532}{{\tt 1608.02532}}].

\bibitem{Christensen:1977jc}
S.~M. Christensen and S.~A. Fulling, \emph{{Trace Anomalies and the Hawking
  Effect}}, \href{http://dx.doi.org/10.1103/PhysRevD.15.2088}{\emph{Phys. Rev.}
  {\bf D15} (1977) 2088--2104}.

\bibitem{Eune:2015xvx}
M.~Eune, Y.~Gim and W.~Kim, \emph{{Effective Tolman temperature induced by
  trace anomaly}},
  \href{http://dx.doi.org/10.1140/epjc/s10052-017-4812-y}{\emph{Eur. Phys. J.}
  {\bf C77} (2017) 244}, [\href{http://arxiv.org/abs/1511.09135}{{\tt
  1511.09135}}].

\bibitem{Eune:2017iab}
M.~Eune and W.~Kim, \emph{{Proper temperature of the Schwarzschild AdS black
  hole revisited}},
  \href{http://dx.doi.org/10.1016/j.physletb.2017.08.009}{\emph{Phys. Lett.}
  {\bf B773} (2017) 57--61}, [\href{http://arxiv.org/abs/1703.00589}{{\tt
  1703.00589}}].

\bibitem{Kim:2017aag}
W.~Kim, \emph{{The effective Tolman temperature in curved spacetimes}},
  \href{http://dx.doi.org/10.1142/S0218271817300257}{\emph{Int. J. Mod. Phys.}
  {\bf D26} (2017) 1730025}, [\href{http://arxiv.org/abs/1709.02537}{{\tt
  1709.02537}}].

\bibitem{Deser:1997ri}
S.~Deser and O.~Levin, \emph{{Accelerated detectors and temperature in
  (anti)-de Sitter spaces}},
  \href{http://dx.doi.org/10.1088/0264-9381/14/9/003}{\emph{Class. Quant.
  Grav.} {\bf 14} (1997) L163--L168},
  [\href{http://arxiv.org/abs/gr-qc/9706018}{{\tt gr-qc/9706018}}].

\bibitem{Deser:1998xb}
S.~Deser and O.~Levin, \emph{{Mapping Hawking into Unruh thermal properties}},
  \href{http://dx.doi.org/10.1103/PhysRevD.59.064004}{\emph{Phys. Rev.} {\bf
  D59} (1999) 064004}, [\href{http://arxiv.org/abs/hep-th/9809159}{{\tt
  hep-th/9809159}}].

\bibitem{Brynjolfsson:2008uc}
E.~J. Brynjolfsson and L.~Thorlacius, \emph{{Taking the Temperature of a Black
  Hole}}, \href{http://dx.doi.org/10.1088/1126-6708/2008/09/066}{\emph{JHEP}
  {\bf 09} (2008) 066}, [\href{http://arxiv.org/abs/0805.1876}{{\tt
  0805.1876}}].

\bibitem{Teitelboim:1983ux}
C.~Teitelboim, \emph{{Gravitation and Hamiltonian Structure in Two Space-Time
  Dimensions}},
  \href{http://dx.doi.org/10.1016/0370-2693(83)90012-6}{\emph{Phys. Lett.} {\bf
  126B} (1983) 41--45}.

\bibitem{Jackiw:1984je}
R.~Jackiw, \emph{{Lower Dimensional Gravity}},
  \href{http://dx.doi.org/10.1016/0550-3213(85)90448-1}{\emph{Nucl. Phys.} {\bf
  B252} (1985) 343--356}.

\bibitem{Spradlin:1999bn}
M.~Spradlin and A.~Strominger, \emph{{Vacuum states for AdS(2) black holes}},
  \href{http://dx.doi.org/10.1088/1126-6708/1999/11/021}{\emph{JHEP} {\bf 11}
  (1999) 021}, [\href{http://arxiv.org/abs/hep-th/9904143}{{\tt
  hep-th/9904143}}].

\bibitem{Mann:1989gh}
R.~B. Mann, A.~Shiekh and L.~Tarasov, \emph{{Classical and Quantum Properties
  of Two-dimensional Black Holes}},
  \href{http://dx.doi.org/10.1016/0550-3213(90)90265-F}{\emph{Nucl. Phys.} {\bf
  B341} (1990) 134--154}.

\bibitem{Muta:1992xw}
T.~Muta and S.~D. Odintsov, \emph{{Two-dimensional higher derivative quantum
  gravity with constant curvature constraint}},
  \href{http://dx.doi.org/10.1143/PTP.90.247}{\emph{Prog. Theor. Phys.} {\bf
  90} (1993) 247--255}.

\bibitem{Lemos:1993qn}
J.~P.~S. Lemos and P.~M. Sa, \emph{{Nonsingular constant curvature
  two-dimensional black hole}},
  \href{http://dx.doi.org/10.1142/S0217732394000587}{\emph{Mod. Phys. Lett.}
  {\bf A9} (1994) 771--774}, [\href{http://arxiv.org/abs/gr-qc/9309023}{{\tt
  gr-qc/9309023}}].

\bibitem{Kumar:1994ve}
A.~Kumar and K.~Ray, \emph{{Thermodynamics of two-dimensional black holes}},
  \href{http://dx.doi.org/10.1016/0370-2693(95)00374-T}{\emph{Phys. Lett.} {\bf
  B351} (1995) 431--438}, [\href{http://arxiv.org/abs/hep-th/9410068}{{\tt
  hep-th/9410068}}].

\bibitem{Lemos:1996bq}
J.~P. Lemos, \emph{{Thermodynamics of the two-dimensional black hole in the
  Teitelboim-Jackiw theory}},
  \href{http://dx.doi.org/10.1103/PhysRevD.54.6206}{\emph{Phys.\ Rev.} {\bf
  D54} (1996) 6206--6212}, [\href{http://arxiv.org/abs/gr-qc/9608016}{{\tt
  gr-qc/9608016}}].

\bibitem{Polyakov:1987zb}
A.~M. Polyakov, \emph{{Quantum Gravity in Two-Dimensions}},
  \href{http://dx.doi.org/10.1142/S0217732387001130}{\emph{Mod. Phys. Lett.}
  {\bf A2} (1987) 893}.

\bibitem{Bilal:1992kv}
A.~Bilal and C.~G. Callan, Jr., \emph{{Liouville models of black hole
  evaporation}},
  \href{http://dx.doi.org/10.1016/0550-3213(93)90102-U}{\emph{Nucl. Phys.} {\bf
  B394} (1993) 73--100}, [\href{http://arxiv.org/abs/hep-th/9205089}{{\tt
  hep-th/9205089}}].

\bibitem{NavarroSalas:1995vd}
J.~Navarro-Salas, M.~Navarro and C.~F. Talavera, \emph{{Weyl invariance and
  black hole evaporation}},
  \href{http://dx.doi.org/10.1016/0370-2693(95)00848-F}{\emph{Phys. Lett.} {\bf
  B356} (1995) 217--222}, [\href{http://arxiv.org/abs/hep-th/9505139}{{\tt
  hep-th/9505139}}].

\bibitem{Cadoni:1994uf}
M.~Cadoni and S.~Mignemi, \emph{{Nonsingular four-dimensional black holes and
  the Jackiw-Teitelboim theory}},
  \href{http://dx.doi.org/10.1103/PhysRevD.51.4319}{\emph{Phys. Rev.} {\bf D51}
  (1995) 4319--4329}, [\href{http://arxiv.org/abs/hep-th/9410041}{{\tt
  hep-th/9410041}}].

\bibitem{Kim:1995wr}
W.~T. Kim and J.~Lee, \emph{{Hawking radiation and energy conservation in an
  evaporating black hole}},
  \href{http://dx.doi.org/10.1103/PhysRevD.52.2232}{\emph{Phys. Rev.} {\bf D52}
  (1995) 2232--2238}, [\href{http://arxiv.org/abs/hep-th/9502115}{{\tt
  hep-th/9502115}}].

\bibitem{Callan:1992rs}
C.~G. Callan, Jr., S.~B. Giddings, J.~A. Harvey and A.~Strominger,
  \emph{{Evanescent black holes}},
  \href{http://dx.doi.org/10.1103/PhysRevD.45.R1005}{\emph{Phys. Rev.} {\bf
  D45} (1992) R1005}, [\href{http://arxiv.org/abs/hep-th/9111056}{{\tt
  hep-th/9111056}}].

\bibitem{Deser:1976yx}
S.~Deser, M.~J. Duff and C.~J. Isham, \emph{{Nonlocal Conformal Anomalies}},
  \href{http://dx.doi.org/10.1016/0550-3213(76)90480-6}{\emph{Nucl. Phys.} {\bf
  B111} (1976) 45--55}.

\bibitem{BoschiFilho:1991xz}
H.~Boschi-Filho and C.~P. Natividade, \emph{{Anomalies in curved space-time at
  finite temperature}},
  \href{http://dx.doi.org/10.1103/PhysRevD.46.5458}{\emph{Phys. Rev.} {\bf D46}
  (1992) 5458--5466}.

\bibitem{Boulware:1974dm}
D.~G. Boulware, \emph{{Quantum Field Theory in Schwarzschild and Rindler
  Spaces}}, \href{http://dx.doi.org/10.1103/PhysRevD.11.1404}{\emph{Phys. Rev.}
  {\bf D11} (1975) 1404}.

\bibitem{Unruh:1977ga}
W.~G. Unruh, \emph{{Origin of the Particles in Black Hole Evaporation}},
  \href{http://dx.doi.org/10.1103/PhysRevD.15.365}{\emph{Phys. Rev.} {\bf D15}
  (1977) 365--369}.

\bibitem{Kim:1998wy}
W.~T. Kim, \emph{{AdS(2) and quantum stability in the CGHS model}},
  \href{http://dx.doi.org/10.1103/PhysRevD.60.024011}{\emph{Phys. Rev.} {\bf
  D60} (1999) 024011}, [\href{http://arxiv.org/abs/hep-th/9810055}{{\tt
  hep-th/9810055}}].

\bibitem{Rocha:2008fe}
J.~V. Rocha, \emph{{Evaporation of large black holes in AdS: Coupling to the
  evaporon}},
  \href{http://dx.doi.org/10.1088/1126-6708/2008/08/075}{\emph{JHEP} {\bf 08}
  (2008) 075}, [\href{http://arxiv.org/abs/0804.0055}{{\tt 0804.0055}}].

\bibitem{Unruh:1983ms}
W.~G. Unruh and R.~M. Wald, \emph{{What happens when an accelerating observer
  detects a Rindler particle}},
  \href{http://dx.doi.org/10.1103/PhysRevD.29.1047}{\emph{Phys. Rev.} {\bf D29}
  (1984) 1047--1056}.

\bibitem{doi:10.1142/p378}
A.~Fabbri and J.~Navarro-Salas, \emph{Modeling Black Hole Evaporation}.
\newblock PUBLISHED BY IMPERIAL COLLEGE PRESS AND DISTRIBUTED BY WORLD
  SCIENTIFIC PUBLISHING CO., 2005.
\newblock 10.1142/p378.

\end{thebibliography}\endgroup

\end{document}